\documentclass[final,5p,times,twocolumn]{elsarticle}

\usepackage{graphicx}

\usepackage{amssymb}
\usepackage{amsthm}

\journal{Physica C: Superconductivity and its Applications~}

\def\pra{Phys. Rev. A}

\def\prl{Phys. Rev. Lett.}
\def\Ek{E_\mathbf{k}}
\def\ek{\epsilon_\mathbf{k}}
\def\sumk{\sum_\mathbf{k}}

\begin{document}

\begin{frontmatter}



\title{Superfluidity in atomic Fermi gases}

\author[label1]{Yi Yu} 
\author[label2]{Qijin Chen\corref{cor1}}
\cortext[cor1]{Corresponding author: qchen@zju.edu.cn}
\address[label1]{Center for Measurement and Analysis, Zhejiang
  University of Technology, 18 Chaowang Rd., Hangzhou 310014, China}
\address[label2]{Zhejiang Institute of Modern Physics and Department of
  Physics, Zhejiang University, 38 Zheda Rd., Hangzhou 310027, China}

\begin{abstract}

  In a trapped atomic Fermi gas, one can tune continuously via a
  Feshbach resonance the effective pairing interaction between fermionic
  atoms from very weak to very strong.  As a consequence, the low
  temperature superfluidity evolves continuously from the BCS type in
  the weak interaction limit to that of Bose-Einstein condensation in
  the strong pairing limit, exhibiting a BCS-BEC crossover. In this
  paper, we review recent experimental progress in atomic Fermi gases
  which elucidates the nature of the superfluid phase as the interaction
  is continuously tuned. Of particular interest is the intermediate or
  crossover regime where the $s$-wave scattering length diverges.  We
  will present an intuitive pairing fluctuation theory, and show that
  this theory is in quantitative agreement with existing experiments in
  cold atomic Fermi gases.

\end{abstract}

\begin{keyword}



Atomic Fermi gases \sep superfluidity \sep BCS-BEC crossover \sep
pairing fluctuations

\PACS 03.75.Hh \sep 03.75.Ss \sep 74.20.-z

\end{keyword}

\end{frontmatter}


\section{Introduction}
\label{sec:Intro}



Ultracold atomic Fermi gases have been a very exciting, rapidly
developing field, which has emerged within the past several years,
bridging condensed matter and atomic, molecular and optical physics
\cite{ourreview}.  Using a Feshbach resonance in a magnetic field, one
can tune the effective pairing interaction strength between fermionic
atoms from very weak to very strong \cite{Milstein}. As the interaction
strength varies, the nature of the low temperature superfluidity of
these Fermi gases evolves continuously from the BCS type in the weak
coupling limit to Bose-Einstein condensation (BEC) in the strong pairing
limit, exhibiting a BCS-BEC crossover, which has been of great
theoretical interest since 1960's
\cite{Eagles,Leggett,NSR,ourreview}. Of particular interest is the
unitary regime, where the $s$-wave scattering length diverges. This is a
strongly correlated regime where no small parameter is available for
perturbative expansions. It has been expected that this regime provides
a prototype for studying both high $T_c$ superconductors and strongly
interacting Fermi gases which are also of interest to nuclear and 
astro-physicists.

In this paper, we first review experimental progress in atomic Fermi
gases, with an emphasis on recent radio frequency spectroscopy
measurements. Then we will present a pairing fluctuation theory
compare with experiment. We show that this theory successfully explain
experimental measurements.

\section{Experimental progress}

The first theoretical study of BCS-BEC crossover dates back to 1960's,
although it did not get much attention until the seminal work of Leggett
in 1980 on BCS-BEC crossover at zero temperature \cite{Leggett}. The
discovery of high Tc superconductivity in 1986 gave a strong boost to
the interest in BCS-BEC crossover
\cite{Uemura,TDLee1,Randeria,Janko,Chen2,ourreview}. It was suggested
that the unusual pseudogap phenomena in the cuprate superconductors might
have to do with BCS-BEC crossover.  Experimental efforts in this area
fell far behind, because it had been difficult to find a system where
the attractive pairing interaction is tunable. Thanks to the laser
cooling and trapping technique in 1990's, one is able to create
``artificial'' many-body systems of fermionic atoms in a laboratory. The
existence of a Feshbach resonance in these Fermi gases makes it possible
to tune the interaction strength.

For ease of control, the Feshbach resonances for the two widely studied
species, $^6$Li and $^{40}$K, are both very wide.  The interaction in
both cases are of the short-range, $s$-wave type.
 They are often taken to be a contact potential in theoretical
 treatments.

The first experimental realization of BCS-BEC crossover was achieved in
2004 by Jin and coworkers, \cite{Jin4,Jin_us} and almost the same time
by the Grimm group \cite{Grimm4} and the Ketterle group
\cite{Ketterle3}. Due to the difficulty in tuning temperature $T$, the
Fermi gases were either in the superfluid or normal state at given
interaction strength (or the magnetic detuning). Continuous variation of
the system as a function of temperature was first realized by the Thomas
group \cite{Kinast} at unitarity. In collaboration with the theory group
at Chicago \cite{ChenThermo}, Thomas \textit{et al} \cite{ThermoScience}
observed for the first time continuous phase transition from the normal
to superfluid state in a unitary $^6$Li gas. One could argue, of course,
that the vortex measurement of the Ketterle group provided the most
definitive smoking gun for a superfluid state. \cite{KetterleV}

Besides the interaction strength, another great tunability is population
imbalance between the two fermionic species to be paired
\cite{Stability}. It adds a whole new dimension to the phase diagram and
makes the physics much richer. It also generates interest \cite{LOFF1} in
possible observation of the Larkin-Ovchinnikov-Fulde-Ferrell (LOFF)
state \cite{FFLO}. Experimental work in population imbalanced Fermi
gases was pioneered by the Hulet group \cite{Rice1} and the Ketterle
group \cite{Zwierlein2006}.  
 Experiment in the extreme population imbalanced limit by the Ketterle
 group manifested \cite{Schunck07} the importance of Hartree-like
 correlation effects besides BCS-type of pairing.

Unlike an electron system, it has been difficult to measure the
excitation gap in the Fermi gas superfluid. Among all experimental
techniques, Radio frequency (RF) spectroscopy \cite{Grimm4} is arguably
the most direct probe. Using a tunable RF field to excite one of the two
pairing atoms from a lower hyperfine state (level 2) to a higher
hyperfine level 3 which do not participate in pairing, a higher
frequency will be needed if the atoms in level 2 are paired. Such a
frequency shift (detuning) provides a good measure of the excitation
gap. Previous measurement by Grimm and coworkers \cite{Grimm4},
 and later repeated by the Ketterle group \cite{Schunck07},
was done in a momentum integrated fashion.  At low $T$, the RF spectra
displayed double-peak structure, with a sharp peak at zero detuning and
a broad peak at positive detuning.  This double-peak feature was nicely
interpreted \cite{Torma2,heyan} as transitions from unpaired atoms the
trap edge (corresponding to the sharp peak) and from a distribution of
paired atoms in the inner part of the trap (broad peak). However, doubt
was cast about the origin of the two peaks as to whether they reflect
pairing of bound state effects \cite{Schunck07} or simply a result of
trap inhomogeneity \cite{Mueller07}. Recently, attention was also drawn
to final state effects both theoretically \cite{Basu,ourRF3} and
experimentally \cite{Schunck07}.

A big step in the RF technique was the recent momentum-resolved RF
spectroscopy experiment in $^{40}$K by the Jin group \cite{Jin6}. With
momentum resolution, RF spectroscopy is equivalent to the angle-resolved
photoemission spectroscopy (ARPES) for an electron system, 
 In fact, it is cleaner than ARPES in that ARPES is only a
 two-dimensional probe, which is often plagued by the existence of
 surface states, surface contaminations, work function, and the
 complication of energy dispersion in the third dimension. In
 comparison, of course, the signal-to-noise ratio in a Fermi gas
 experiment is much lower, as limited by the (low) total number of
 atoms in the gas.
although the trap inhomogeneity adds complication to the interpretation
of the spectrum.  Like ARPES, momentum-resolved RF spectroscopy measures
the fermion spectral function, $A(\mathbf{k},\omega)$, which is of
central importance in characterizing the system. 

\section{Theoretical Formalism}

In this section, we now present a simple pairing fluctuation theory,
which was first developed \cite{Chen2} to explain the pseudogap
phenomena in high Tc superconductors.
Fermi gases in the presence of a Feshbach resonance can be effectively
described by a two-channel model \cite{Milstein}. It has now been known
that the closed-channel fraction \cite{Hulet4,ChenClosed} is very small
for both $^6$Li and $^{40}$K, throughout the BCS-BEC
crossover. Therefore, for these systems, a one-channel model is often
used as a good approximation, given
by the grand canonical Hamiltonian
\begin{eqnarray}
 \lefteqn{\hspace*{-2ex}H - \sum_\sigma \mu_\sigma N_\sigma =
   \sum_{\mathbf{k},\sigma} 
(\epsilon_\mathbf{k}^{} - \mu_\sigma) a_{\mathbf{k},\sigma}^\dag
a_{\mathbf{k},\sigma}^{}}& \nonumber\\
&\hspace*{-2ex}+{} \sum_{\bf q,k,k'} U(\mathbf{k,k'})
a^\dag_{\mathbf{q/2+k},\uparrow} a^\dag_{\mathbf{q/2-k},\downarrow}
a^{}_{\mathbf{q/2-k'},\downarrow} a^{}_{\mathbf{q/2+k'},\uparrow},
\label{eq:Hamiltonian}
\end{eqnarray} 
where $\epsilon^{}_\mathbf{k} = \hbar^2k^2/2m$ is the free atom
dispersion.  The difference between Eq.~(\ref{eq:Hamiltonian}) and its
BCS counterpart is that BCS keeps only the $\mathbf{q} = 0$ term in the
interactions.  The inclusion of finite $\mathbf{q}$ terms allows
incoherent, finite momentum pairing.  For clarity of presentation, we
will take a contact potential, $ U(\mathbf{k,k'}) = 1$, and use a
4-momentum notation, $K=(\mathbf{k},i\omega_n)$,
$Q=(\mathbf{q},i\Omega_l)$, $ \sum_K = T \sum_\mathbf{k} \sum_{n}$, and
set $\hbar=1$. Population imbalance can be described by $\mu_\uparrow
\ne \mu_\downarrow$. However, here we will only present the equations
for the case of equal spin mixture. Generalization to population
imbalance can be found in Ref.~\cite{Chien06}.

We assume that (i) the fermionic self energy $\Sigma$ has a pairing
origin, (ii) pairs can be either condensed or fluctuating with a finite
momentum, and (iii) condensed and noncondensed pairs do not mix at the
level of $T$-matrix approximation. Figure 1 shows diagrammatically the
contributions to the self-energy, where the double (red) lines indicate
finite momentum pairs and the dotted line indicates the condensate. The
subscripts ``sc'' and ``pg'' stand for superfluid condensate and
pseudogap contributions, respectively.

\begin{figure}
\centerline{\includegraphics[width=3.2in,clip]{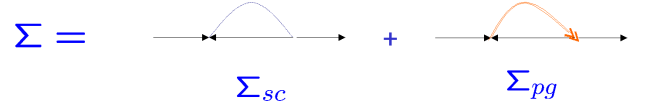}}
\caption{Schematic diagrams for the fermionic self-energy
  $\Sigma(K)$. The dotted and (red) double lines represent the
  condensate and finite momentum pairs, respectively.}
\end{figure}

To tackle this problem, we use a Green's function method. We derive the
equations of motion  for one- and two-particle Green's functions
$G$ and $G_2$,
 which will involve higher order, three particle Green's functions
$G_3$:
$i \dot G = [H, G] \sim G, G_2$, $i \dot G_2 = [H, G_2] \sim G, G_2,
G_3$. We then truncate the equations of motion at the level of $G_3$,
factorize $G_3$ into a sum of products of $G$ and $G_2$, and treat $G$
and $G_2$ on equal footing. 
For $G_2$, we focus on the particle-particle channel, neglecting
the particle-hole channel which normally only provides a
chemical potential shift. 
 We emphasize that
\emph{it is the particle-particle channel that gives rise to
  superfluidity.} After some lengthy but straightforward derivation, we
obtain the self energy:
\begin{eqnarray}
\Sigma(K) &=& \Sigma_{sc}(K) + \Sigma_{pg} (K),\\
\Sigma_{sc}(K)  &=& -\Delta_{sc}^2G_0(-K)
,\\
\Sigma_{pg}(K)  &=&  \sum_Q t_{pg}(Q) G_0(Q-K), 
\end{eqnarray}
where 
\begin{equation} 
t_{pg}(Q) = \sum_Q \frac{U}{1+U\chi(Q)} 
\end{equation}
 is the (pseudogap)
$T$-matrix, and $\chi(Q) = \sum_K G_0(Q-K) G(K) $ is the pair
susceptibility. Here $G_0$ is the bare Green's function. A detailed
derivation of this result can be found in Ref.~\cite{ChenPhD}. Note that
the $T$-matrix is effectively a renormalized pairing interaction. It
shares exactly the same pole structure as the two-particle Green's
function, $G_2$. Through a Taylor expansion of its denominator, one can
extract the pair dispersion: 
\begin{equation}
t_{pg}^{-1}(Q) \approx
Z(i\Omega_l-\Omega_q +\mu_{pair})\,.
\end{equation}
 The superfluid instability is given
by $1+U\chi(0) = 0 \propto \mu_{pair}$, which is the BEC condition for
pairs. Note that $\chi(Q)$ involves a mix of bare and full Green's
functions. We emphasize that \emph{this is a natural consequence of the
  equation of motion technique} since it involves the operator
$\hat{G}_0^{-1}$. \emph{It is this $G_0G$ form of $\chi$ that leads back to
  the BCS-form of gap equation in the absence of finite momentum pairs.}

We focus on the superfluid phase where $t_{pg}(Q)$ diverges at
$Q=0$. Defining 
\begin{equation}
\Delta_{pg}^2 \equiv -\sum_{Q\ne0} t_{pg}(Q), 
\label{eq:pg}
\end{equation}
we have
\begin{eqnarray}
\Sigma_{pg}(K) &=& -\left[\sum_Q t_{pg}(Q)\right]G_0(-K) +\delta
\Sigma \nonumber\\ 
&=&{} -\Delta_{pg}^2 G_0(-K) +\delta \Sigma .
\end{eqnarray}
Neglecting the residue term $\delta\Sigma$, $\Sigma_{pg}$ takes the same
form as $\Sigma_{sc}$. Thus we have immediately the BCS form of total
self energy, $\Sigma(K) = -\Delta^2 G_0(-K)$, with $\Delta^2 =
\Delta_{sc}^2 + \Delta_{pg}^2$. This then leads to the BCS form of gap
equation,
\begin{equation}
1+U\sum_\mathbf{k} \frac{1-2f(\Ek)}{2\Ek} = 0, 
\label{eq:gap}
\end{equation}
where $ \Ek = \sqrt{(\ek-\mu)^2+\Delta^2} $ is the quasiparticle
dispersion. \emph{Different from the BCS mean-field theory, we emphasize
  that here $\Delta^2 $ contains contributions from both condensed and
  noncondensed pairs so that it in general does not vanish at $T_c$.}
Note that the finite $\mathbf{q}$ pairs are different from the order
parameter collective modes; the latter represent a coherent motion of
the condensate.  Here $\Delta_{sc}^2$ and $\Delta_{pg}^2$ are loosely
proportional to the density of condensed and noncondensed pairs,
respectively.

\begin{figure}
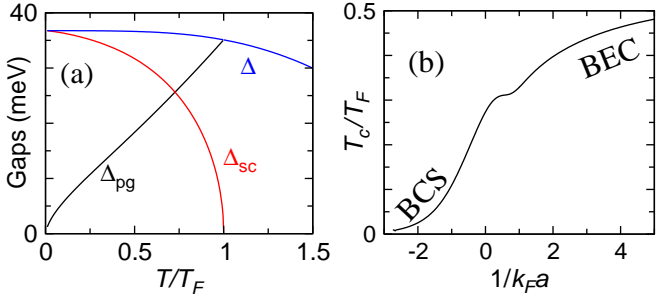

\centerline{\includegraphics[width=1.7in,clip]{Fig2a.eps}
\includegraphics[width=1.65in,clip]{Fig2b.eps}}
\caption{Typical behavior of (a) the temperature dependence of the gaps
  in a pseudogapped superfluid, and of (b) $T_c$ as a function of
  $1/k_Fa$ in a trap, where $k_F$ is the noninteracting Fermi momentum,
  and $a$ is the $s$-wave scattering length.}
\end{figure}

Equations (\ref{eq:gap}) and (\ref{eq:pg}), along with the number equation
\begin{equation}
n = 2\sum_K G(K),
\label{eq:number}
\end{equation}
form a closed set of equations for the homogeneous case, which can be
used to solve for $\mu$, $T_c$, and the gaps at $T\le T_c$. $T_c$ is
determined by setting $\Delta_{sc}=0$. Typical behaviors of the gaps are
shown in Fig.~2(a).

To address Fermi gases in a trap, we use the local density
approximation, by replacing $\mu \rightarrow \mu - V_{trap}(r)$. Then
the number equation becomes $N = \int d^3 r \, n(r)$. In Fig.~2(b) we
show the BCS-BEC crossover behavior of $T_c$ in a trap. Here $1/k_Fa$
parametrizes the interaction strength.


The RF response can be derived using the linear response theory. The RF
interaction is described by 
\begin{equation}
H_{rf} = e^{i\Omega t}\int d^3x\,\psi_3^\dag\psi_2^{} +h.c., 
\end{equation}
and the response Kernel by 
\begin{equation}
D(i\Omega_l)
= \sum_K G^{(2)}(K)G^{(3)}(K+Q).
\end{equation}
  We assume hyperfine level 3 was
initially empty. In the absence of final state interactions, as in
$^{40}$K, we obtain \cite{ChenPRL2009} the RF current 
\begin{eqnarray}
I(k,\nu)
&=&-\frac{1}{\pi} \mbox{Im}\, D^R(\nu+\mu-\mu_3)\nonumber\\
 &=&\left. \frac{1}{2\pi}\sumk A(\mathbf{k},\omega)f(\omega)
\right|_{\omega=\ek-\mu-\nu}.
\end{eqnarray}
 In order to address $A(\mathbf{k},\omega)
= -2\, \mbox{Im}\, G(\mathbf{k}, \omega+i 0^+)$ properly, we need to
include the lifetime effects of finite momentum pairs and add an
incoherent term $i\Sigma_0$ in (and only in) $\Sigma_{pg}$, reflecting
the residue term $\delta\Sigma$ which we drop in solving the set of
equations, i.e., 
\begin{equation}
\Sigma_{pg}(\mathbf{k},\omega) =
\frac{\Delta^2_{pg}}{\omega+\ek-\mu+i\gamma} -i\Sigma_0. 
\end{equation}
While above $T_c$
the spectral function with a pseudogap constitute a double peak
structure with suppressed spectral weight at the Fermi level, below
$T_c$, there is a zero at $\omega = -(\ek -\mu)$. As $\Delta_{sc}$
increases with decreasing $T$ below $T_c$, the spectral peaks sharpen
rapidly. This is a phase coherence effect. The parameters $\gamma$ and
$\Sigma_0$ can be estimated from experimental RF spectra.

\begin{figure}
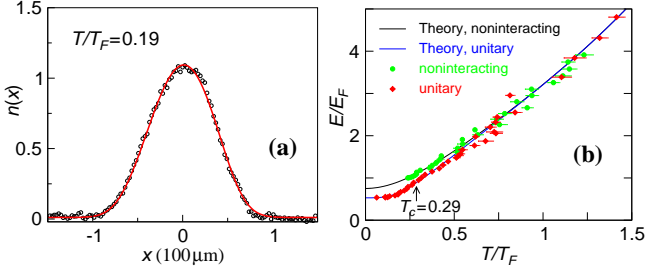

\centerline{\includegraphics[width=1.65in,clip]{Fig3a.eps}
\includegraphics[width=1.65in,clip]{Fig3b.eps}}
\caption{Comparison of (a) density profile and (b) energy $E/E_F$ for a
  unitary  $^6$Li gas between theory (curves) and experiment
  (symbols). Also shown in (b) is comparison for the noninteracting
  energy. Here $E_F=k_BT_F$ is the noninteracting Fermi energy.}
\end{figure}

\section{Comparison between theory and experiment}

In Fig.~3, we compare between theory (curves) and experiment (symbols)
(a) the density profile \cite{JS5} and (b) system energy
\cite{ThermoScience} for $^6$Li in the unitary limit.  Both experimental
and theoretical density profiles are very smooth, in good agreement with
each other. Alternative theories predicts either a kink at the edge of
the superfluid core or nonmonotonic radial and temperature
dependence. The energy comparison also reveals a quantitative
agreement. The fact that the unitary and noninteracting curves merge at
$T\approx 0.6T_F \gg T_c$ manifests the presence of a pseudogap. It
should be noted that \emph{there is no fitting parameter in our theoretical
  calculations.}

Shown in Fig.~4 is a comparison of the spectral intensity map as a
function of $k$ and single-particle energy $\omega+\mu$ between
experiment \cite{Jin6} and theory \cite{ChenPRL2009} for a unitary
$^{40}$K gas at a temperature slightly above $T_c$. The white dashed
curve is the experimentally extracted quasiparticle dispersion, whereas
the solid curve is obtained theoretically following the experimental
procedure. It is evident that theoretical and experimental results are
rather close to each other. Indeed, as $T$ decreases from above to below
$T_c$, the spectral intensity map evolves \cite{ChenPRL2009} from an
upward dispersing branch at high $T$ to a bifurcation around $T_c$, and
finally to a downward dispersing branch at $T\ll T_c$.  This result
establishes the actual single particle dispersions which contribute to
the RF current, revealing that the broad peak in previous
momentum-integrated RF spectra \cite{Grimm4} indeed has a pairing
origin. Furthermore, it also shows that, despite the trap inhomogeneity,
momentum resolved RF spectroscopy can still provide a quantitative
measure of the spectral function and single particle dispersion. It also
lends support for the present $G_0G$ scheme since alternative NSR-based
theories do not \cite{Muellerprivate} seem to generate the
two-branch-like feature observed in Ref.~\cite{Jin6}. The downward
dispersion around (and above) $T_c$ provides direct evidence for the
existence of a pseudogap above $T_c$ at unitarity. Our theory serves as
a basis for momentum-resolved RF spectroscopy analysis.

In summary, we have presented a pairing fluctuation theory where finite
momentum pairing plays a progressively more important role as the
pairing strength increases, leading to a pseudogap in the single
particle excitation spectrum. This theory has been successfully applied
to multiple experiments in atomic Fermi gases.

This work was supported by Zhejiang University and NSF of China Grant
No.~10974173.

\begin{figure}
\centerline{\includegraphics[width=1.65in,clip]{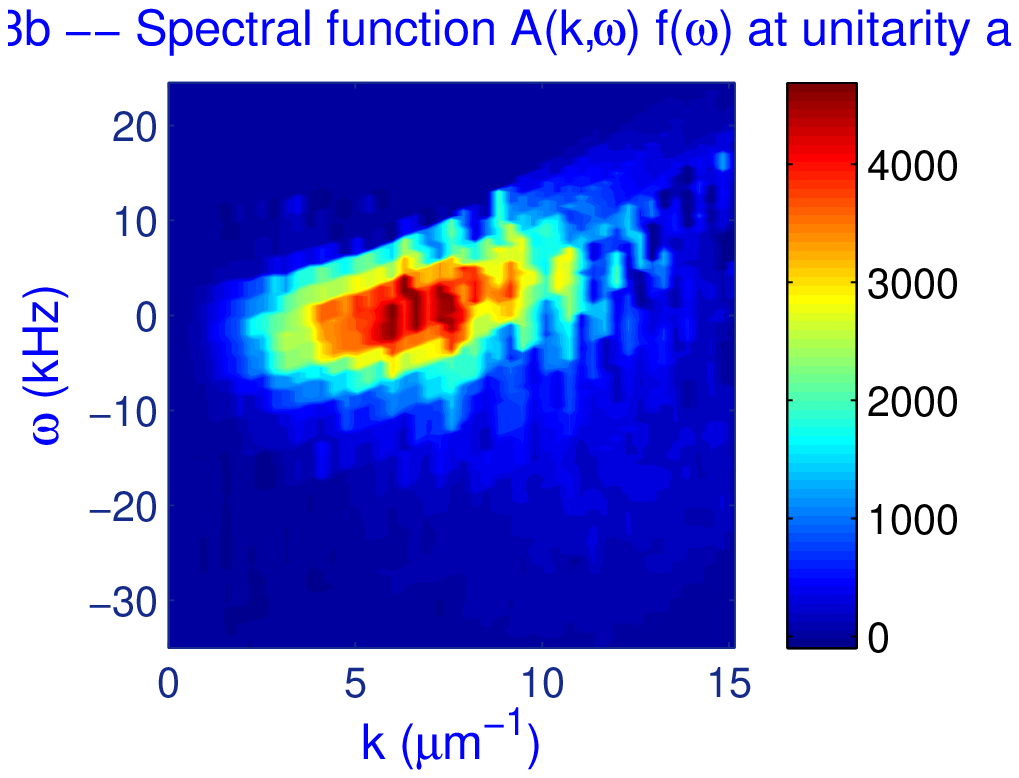}
\includegraphics[width=1.64in,clip]{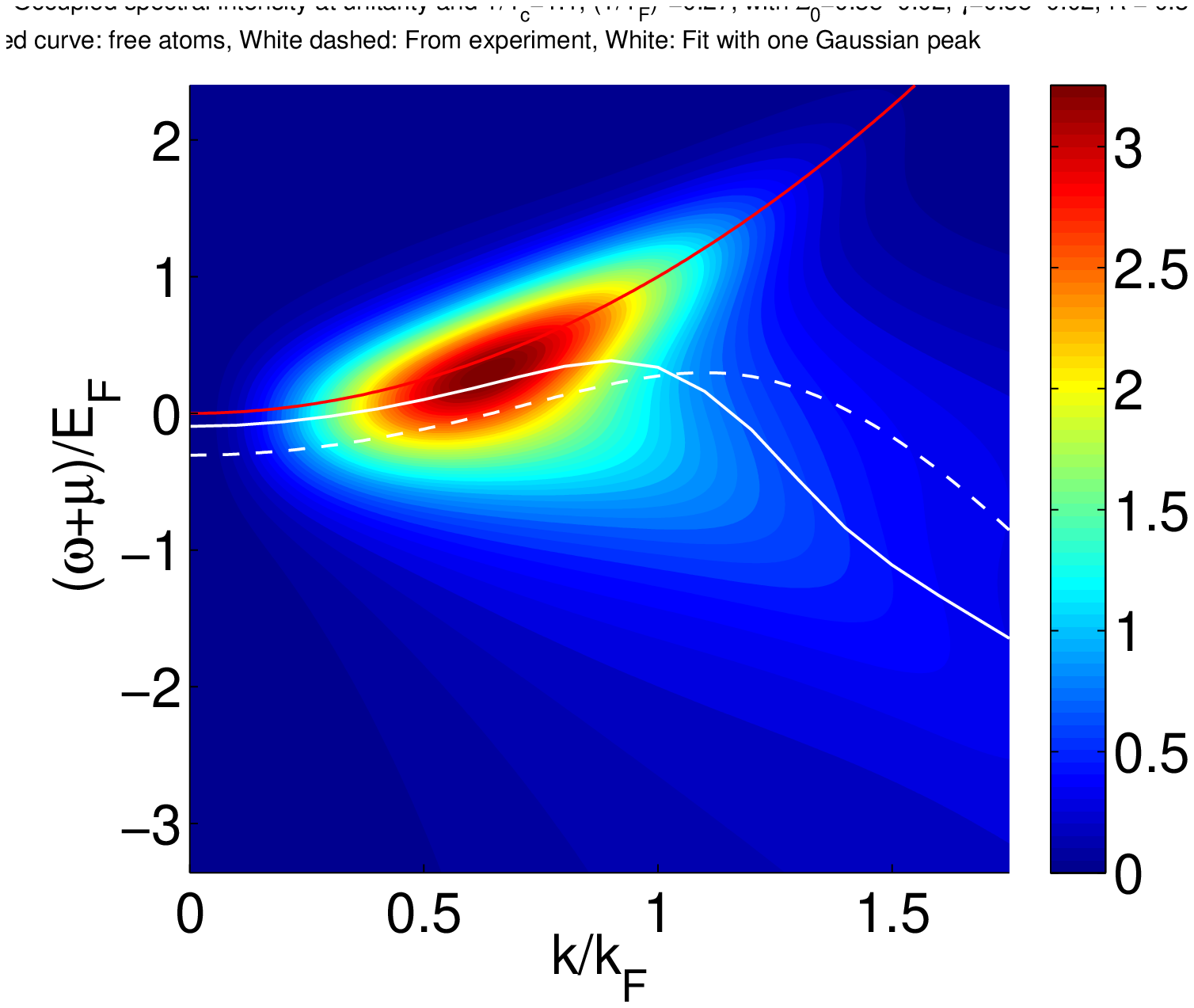} }
\caption{Comparison of spectral intensity map $I(k,\nu)k^2/(2\pi^2)$
  between experiment (left) from Ref.~\cite{Jin6} and theory
  (right). The white dashed curve is an experimental extracted
  quasiparticle dispersion, and the white solid line is obtained
  theoretically following the same experimental data analysis procedure.}
\end{figure}


\end{document}